\documentclass[a4paper,10pt,english]{article}
\usepackage{amsmath}
\usepackage{graphicx}
\usepackage{float}
\usepackage{caption}

\title{Mechanoradicals in tensed tendon collagen as a new source of oxidative stress}

\date{}

\begin{document}

\maketitle

\noindent
\author{Christopher Zapp$^{1,2}$, Agnieszka Obarska-Kosinska$^{1,3}$, Benedikt Rennekamp$^{1,2}$,\\
Davide Mercadante$^{4}$, Uladzimir Barayeu$^{5,6}$, Tobias P. Dick$^{6}$,\\
Vasyl Denysenkov$^{7}$, Thomas Prisner$^{7}$, Marina Bennati$^{8}$, Csaba Daday$^{1,9}$,\\
Reinhard Kappl$^{10}$~\& Frauke Gr\"ater$^{1,9,\dagger}$\\
\\
\normalsize{$^{1}$Heidelberg Institute for Theoretical Studies,}\\
\normalsize{Schloss-Wolfsbrunnenweg 35, 69118 Heidelberg, Germany}\\ 
\normalsize{$^{2}$Institute for Theoretical Physics, Heidelberg University,}\\ \normalsize{Philosophenweg 16,  69120 Heidelberg, Germany}\\ 
\normalsize{$^{3}$Hamburg Unit c/o DESY, European Molecular Biology Laboratory,}\\
\normalsize{Notkestrasse 85, 22607, Hamburg, Germany}\\ 
\normalsize{$^{4}$Biochemical Institute, University of Zuerich,}\\
\normalsize{Winterthurerstr. 190, 8057 Zuerich, Switzerland}\\ 
\normalsize{$^{5}$Faculty of Biosciences, Heidelberg University,}\\
\normalsize{Im Neuenheimer Feld 234, 69120 Heidelberg, Germany}\\ 
\normalsize{$^{6}$Division of Redox Regulation, DKFZ-ZMBH Alliance,}\\
\normalsize{German Cancer Research Center (DKFZ), Im Neuenheimer Feld 280, 69120 Heidelberg, Germany}\\ 
\normalsize{$^{7}$Institute of Physical and Theoretical Chemistry, Goethe University Frankfurt,}\\
\normalsize{Max-von-Laue-Str.~7, 60438 Frankfurt am Main, Germany}\\ 
\normalsize{$^{8}$Max Planck Institute for Biophysical Chemistry,}\\
\normalsize{Am Fassberg 11, 37077 G\"ottingen, Germany}\\ 
\normalsize{$^{9}$Interdisciplinary Center for Scientific Computing, Heidelberg University,}\\
\normalsize{INF 205, 69120 Heidelberg, Germany}\\ 
\normalsize{$^{10}$Institute for Biophysics, Saarland University Medical Center,}\\
\normalsize{CIPMM Geb. 48, 66421 Homburg/Saar, Germany}\\ 
\\
\normalsize{$^\dagger$Corresponding author; E-mail:  frauke.graeter@h-its.org}
}


\section*{Abstract}
As established nearly a century ago, mechanoradicals originate from homolytic bond scission in polymers. The existence, nature and biological relevance of mechanoradicals in proteins, instead, are unknown. We here show that mechanical stress on collagen produces radicals and subsequently reactive oxygen species, essential biological signaling molecules. Electron-paramagnetic resonance (EPR) spectroscopy of stretched rat tail tendon, atomistic Molecular Dynamics simulations and quantum calculations show that the radicals form by bond scission in the direct vicinity of crosslinks in collagen. Radicals migrate to adjacent clusters of aromatic residues and stabilize on oxidized tyrosyl radicals, giving rise to a distinct EPR spectrum consistent with a stable dihydroxyphenylalanine (DOPA) radical. The protein mechanoradicals, as a yet undiscovered source of oxidative stress, finally convert into hydrogen peroxide. Our study suggests collagen I to have evolved as a radical sponge against mechano-oxidative damage and proposes a new mechanism for exercise-induced oxidative stress and redox-mediated pathophysiological processes.



\section*{Introduction}
Polymers subjected to mechanical stress -- be it a shoe sole or rubber band -- generate mechanoradicals by undergoing homolytic bond scission~\cite{Staudinger:1930, Caruso:2009}. Radicals form in polymers even in presence of water, and then readily convert into reactive oxygen species (ROS)~\cite{Fitch:2014,Baytekin:2012}. 
A prime candidate for mechanoradical formation in biological polymers is collagen. 
As the basic material of tendons, cartilage, ligaments and other connective tissues, collagen is under perpetual mechanical load and provides structural and mechanical stability to virtually all human tissues. 
The viscoelastic properties of collagen have been studied in depth~\cite{Fang:2017,Fratzl:2008}, and have been ascribed to a hierarchical deformation mechanism involving straightening and shearing between and within fibrils and triple helices, as well as triple helix unwinding and eventually covalent bond rupture~\cite{Buehler:2006,Depalle:2015,Gautieri:2011,Masic:2015,Misof:1997,Puxkandl:2002,Zitnay:2017,Varma:2016}.
The only evidence for radicaloid bond rupture in biomaterials stems from electron-paramagnetic resonance (EPR) experiments of cut fingernails and milled bone~\cite{Chandra:1987,Symons:1996}.
If and how mechanoradicals form and function in protein materials under physiological, that is, sub-failure levels of loading are fully unknown.

Here we performed tensile tests of fascicles from rat tail tendon and detected radicals originating from mechanical bond scission using EPR spectroscopy. We built an atomistic model of collagen I fibril and identified the regions around crosslinks to carry maximal stresses in Molecular Dynamics pulling simulations. Our joint simulations and experiments show that radicals from primary irreversible bond scission in these crosslink regions migrate to DOPA, which form by  post-translational modifications from phenylalanine and tyrosine residues. These aromatic residues cluster in evolutionarily highly conserved regions. We suggest these clusters of conserved aromatic residues as possible sponges for mechanoradicals, preventing oxidative damage to the tissue. Furthermore, we showed with light absorbance that the DOPA radicals were finally converted into hydrogen peroxide in the presence of water, putting forward a new role of collagen which is to convert mechanical into oxidative stress in connective tissues. 

\section*{Results}

\subsection*{Collagen forms radicals under load}
To detect tension-induced radical formation in collagen, we subjected fascicles dissected from rat tail tendon to constant forces of 5, 10, 15 and then 20\,N, each for 1000\,s (Supplementary Fig.\,1a). These forces correspond to stresses in the 2\,-\,40\,MPa range, depending on the exact fascicle diameter (see Methods). They thus fall into the regime of physiological stresses of tendons, which can exceed 10 times the body weight ($\sim$ 90\,MPa), and also lie well below the rupture stress of fascicles~\cite{Komi:1990}. In this force regime, we recovered the known force-extension behavior of collagen fascicles (Supplementary Fig.\,1b)~\cite{Fang:2017}. Before and in between each constant loading period, we directly measured the presence of radicals in the fascicle by continuous wave X-band electron paramagnetic resonance (EPR) spectroscopy, which is an established method to probe polymer mechanoradicals~\cite{Zhurkov:1964,Beyer:2005}. With increasing constant force, we observed a significant increase in the EPR signal, which was defined as the difference between the minimal and maximal intensity (Fig.\,1a,b). We could reproduce the increase in radicals with the load for different fascicles from the same rat as well as for fascicles from different rats. The increase in the EPR signal was also observed when extending the duration of force application while keeping the force constant (Supplementary Fig.\,1c). We note that collagen tendons show already a small signal (Fig.\,1a, prior to loading), putatively due to the minor load present during extraction (see Methods). Marino \& Becker~\cite{Marino:1969} describe a similar EPR signal (g-factor$\approx$2.007, width$\approx$10\,G) originating from non-stressed tendons, without giving any interpretation. To directly monitor radical production during loading, we devised an EPR setup that allowed to mount the fascicle within the EPR cavity, while being subjected to 3.43\,N of load (Fig.\,1c). Again, within the timescale of tens of minutes, we observed a steady increase in the amount of free radicals during loading. Radicals remained stable over many hours at the measurement conditions (room temperature, ~25\,\% relative humidity). 

\begin{figure}
\includegraphics[scale=0.14]{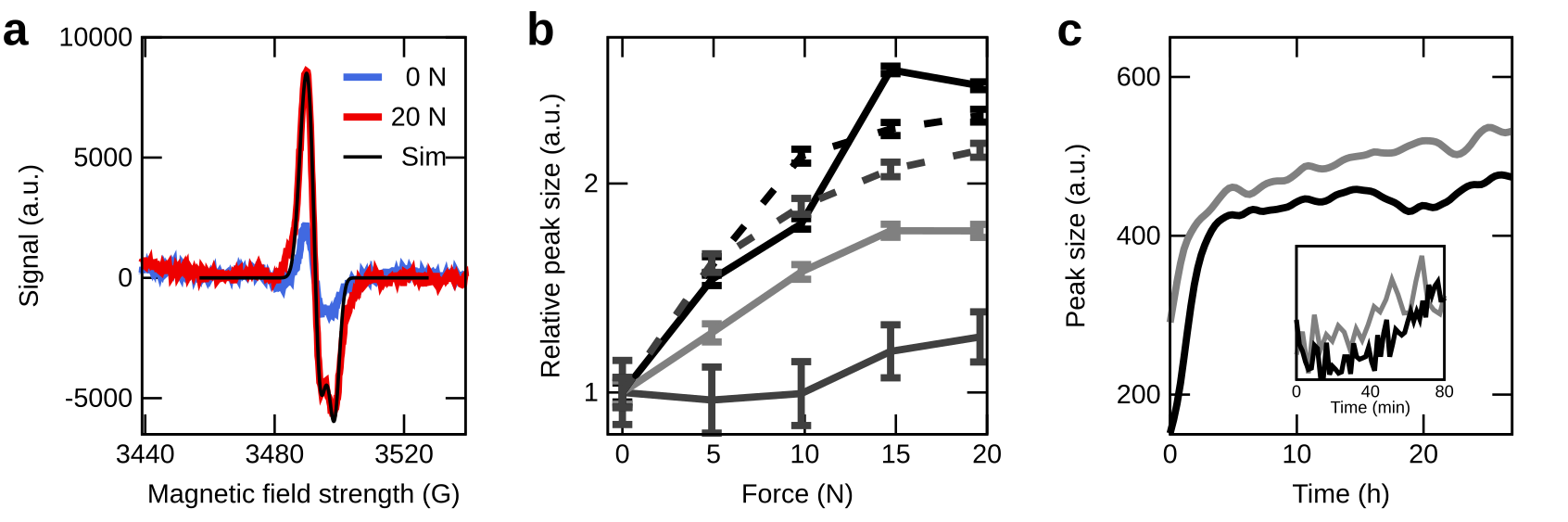} 
\caption{\textbf{Formation of radicals in collagen tendon due to external force.} \textbf{a}, Cw-EPR signal (X-band) against the magnetic field strength measured at room temperature. Blue: unstressed fascicle, red: pulled fascile, black: simulation of pulled fascicle signal.  \textbf{b}, EPR signal size (difference between minimal and maximal intensity) relative to the initial signal before loading for five different samples in different gray tones, continuous and dashed lines for samples pulled at 25\,\% and 45\,\% relative humidity, respectively. Errors show standard deviations of 20 sweeps.  \textbf{c}, Spline-smoothed EPR signal size of two different tendons (black and gray) being pulled at the same time while measuring for 27 hours. The inset shows a zoom for the first 80\,min. of the non-smoothed EPR signal size.}
\end{figure}

\subsection*{Covalent bond rupture is preferred\\ in the vicinity of crosslinks}
To answer if and where covalent bonds rupture and radicals accumulate in collagen materials under subfailure load, we built an atomistic model of a collagen fibril using an integrated structural modeling approach based on a low-resolution fibril structure~\cite{Orgel:2006} and high-resolution collagen-like peptide structures. The resulting model of 67\,nm length spans one overlap and one gap region~\cite{Hodge:1963,Orgel:2006} of a bundle of 37 aligned collagen I triple helices (Fig.\,2a). It contains 12 hydroxylysino-keto-norleucine (HLKNL) crosslinks formed by specific lysine/hydroxylysine side chains~\cite{Eyre:2008} (Fig.\,2b).  
We applied a constant stretching force of 1\,nN per chain to the fibril model in Molecular Dynamics (MD) simulations. We then monitored the distribution of the external force through the fibril structure by Force Distribution Analysis (FDA)~\cite{Costescu:2013}. As expected, the external force propagates along the triple helices and passes over through the crosslinks to adjacent helices (Fig.\,2a). Telopeptides show minimal forces in their backbone bonds as they lie outside of these force propagation pathways. The highest forces concentrate in the N-terminal crosslinks and the backbone of $\sim$10 adjacent residues along the force propagation path (Fig.\,2b,c), suggesting bonds in the $\sim$4\,nm wide cross-sections around these crosslinks to be primary regions of bond rupture and radical formation. C-crosslinks show a less pronounced stress concentration and distribute stress more widely into the gap region (Fig.\,2b,c). Under heterogeneous loading of the collagen triple helices, the dominant loading scenario in a not fully flawless natural tissue, we detected maximal stresses in both N- and C-terminal crosslinks and in their direct vicinity  (Supplementary Fig.\,2b). Taken together, crosslinks and the backbone in their direct vicinity are most stressed.
Therein, the candidates for bond scission are the single bonds, which are of either C-C or C-N type. These two bond types exhibit dissociation energies in the same range, 352-377\,kJ/mol~\cite{Luo:2007}. CASPT2 calculations of representative crosslink and backbone fragments yielded 352-355\,kJ/mol, and confirmed that both C-C and C-N bonds are candidates for mechanoradical formation by homolytic bond scission (Supplementary Fig.\,2e).

\begin{figure}
\includegraphics[scale=0.28]{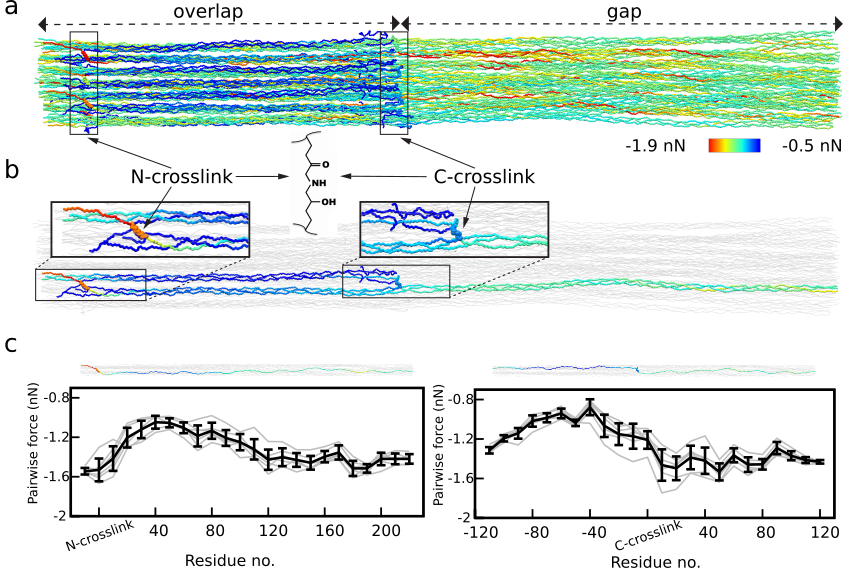}
\caption{\textbf{Force concentrates around crosslinks in tensed collagen.} \textbf{a}, Snapshot of an MD simulation of the crosslinked collagen I model fibril under 1\,nN of external force per chain, colored according to the distribution of the external force through the fibril (blue -- low force, red -- high force). \textbf{b}, Example pair of overlapping triple helices connected by crosslinks from the snapshot shown in \textbf{a}, depicted separately to better visualize forces around the crosslinks. Crosslinks are represented as spheres, the remaining collagen chains as gray ribbons. \textbf{c}, Pairwise forces averaged over 10 consecutive residues along the collagen chains connected by crosslinks (left for N-crosslinks, right for C-crosslinks, with one of the chains shown above the plot aligned to the x-axis) for all six pairs of overlapping triple helices (gray) and average with standard errors over the six pairs (black).}
\end{figure}

\subsection*{Mechanoradicals are stabilized\\ at conserved aromatic residues}
The X-band continuous wave (cw) EPR signal shown in Fig.\,1a is not in line with primary radicals generated from C-N or C-C bond scission, as hyperfine couplings from hydrogen or nitrogen are not resolved, suggesting rapid radical migration through electron transfer reactions. In agreement, mechanically stressing collagen at 77\,K to prevent radical migration yields a distinctly different EPR signal (a typical peroxy signal with a minor contribution from a methylene signal,  suggesting oxygen addition to primary radicals at 77\,K, Supplementary Fig.\,3b,c).
To identify the radical dominant at room temperature, we performed high-frequency G-band (180 GHz) EPR experiments (Fig.\,3a). The obtained signal features a very characteristic axial g-tensor, with g-factors typical for phenoxy-type radicals (Supplementary Table\,1), and no resolved hyperfine coupling, consistent with the X-band EPR spectrum  (Fig.\,1a). 
However, to our knowledge, such an axially symmetric tensor has never been observed in tyrosyl radicals and therefore strongly suggests the formation of another type of phenoxy radical here. An axial g-tensor has been observed by some of us in a dihydroxy-phenylalanine (DOPA) radical~\cite{Nick:2017}, an extra-ordinarily stable tyrosyl radical with an additional hydroxyl group (Fig.\,3b). Interestingly, recent reports on stable DOPA radicals in ribonucleotide reductase reported a rhombic g-tensor~\cite{Srinivas:2018, Blaesi:2018}. We clarified with DFT calculations that the axial g-tensor and the strongly diminished C$_{\beta}$-hydrogen coupling ($<$\,3.6\,G, see Supplementary Table\,1) in stressed collagen originate from a deprotonated DOPA radical anion (Fig.\,3b), i.e. the semiquinone, with high spin density on the two oxygens, instead of the previously reported protonated form with spin density on only one oxygen (Supplementary Table\,1). The quantitative agreement with the DFT computation of DOPA radical anion in isolation implies that hydrogen bonds of DOPA radical anion to neighboring protein residues or water, which are known to shift the g-tensor~\cite{Kaupp:2002}, are either absent or only lead to very minor changes in the signal.

Intriguingly, in the vicinity of crosslinks, the predicted region of bond rupture, collagen I features a strong enrichment in Tyr and Phe (Fig.\,3c,d). Tyr and Phe cluster around crosslinks within the typical 0.4-1.5\,nm distances required for radical transfer (Fig.\,3e,f). Both Tyr and Phe can be oxidized to DOPA by oxygen or superoxide, as shown for various proteins including collagen~\cite{Dean:1993}. They play important roles in radical transfer reactions~\cite{Winkler:2015}, and together (often jointly with Met, Fig.\,3c,d) protect proteins against oxidative damage. 

\begin{figure}
\includegraphics[scale=0.19]{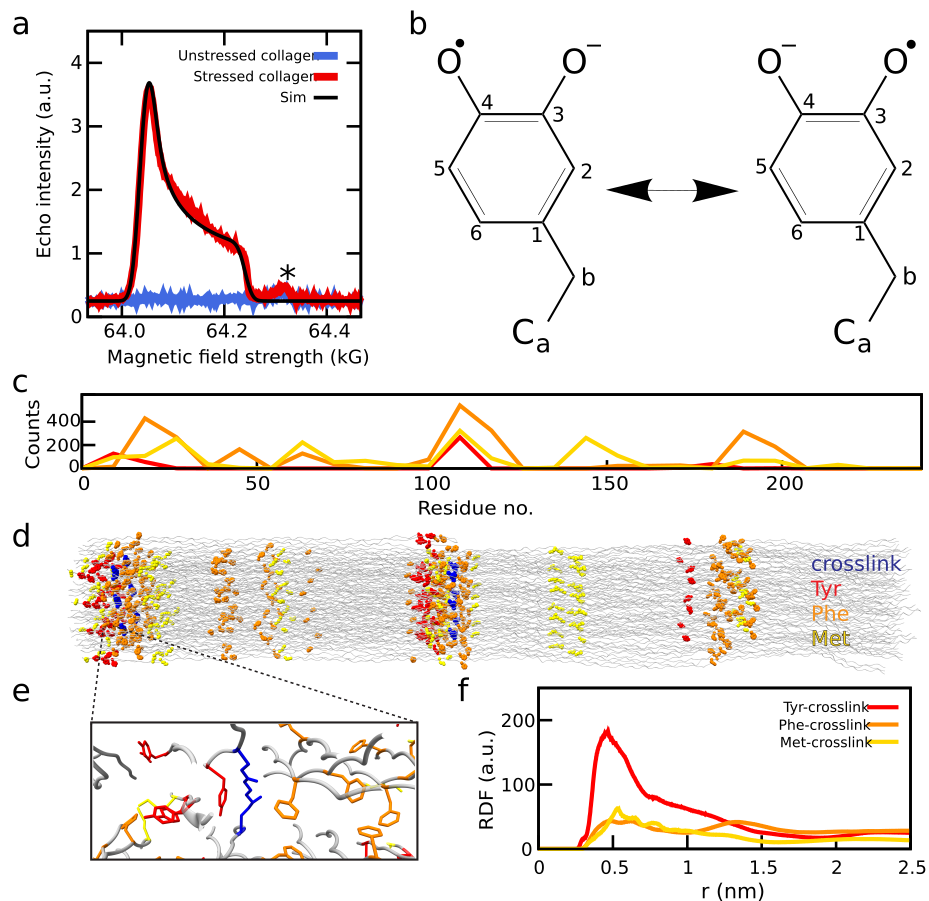}
\caption{\textbf{DOPA acts as radical sink around crosslinks.} \textbf{a}, Echo detected field swept EPR powder pattern (G-band frequencies) at 40\,K of stressed collagen (red) in comparison to unstressed collagen (blue). The simulated powder pattern (for g-tensor values see Supplementary Table\,1) is shown in black. The signal marked by * arises from the quartz capillary. MnO in MgO was used as reference (Supplementary Fig.\,3d). \textbf{b}, DOPA radical anion chemical structure. \textbf{c}, Average counts of the redox-active residues across vertebrates calculated along the fibril from alignments of representative collagen type I sequences (see Methods). The counts were calculated using a moving average of 10 overlapping residues from each chain. \textbf{d}, Distribution of the redox-active residues in the model of the collagen fibril. \textbf{e}, Zoom into crosslink region with high density of redox-active residues. \textbf{f}, Radial distribution function derived from the pulling simulation for redox-active residue-crosslink distances.}
\end{figure}

\subsection*{Mechanoradicals convert into hydrogen peroxide}
In wet tissue, radicals can react with molecular oxygen in water and form reactive oxygen species. 
We used the ferrous oxidation-xylenol orange (FOX) assay to detect hydrogen peroxide in stretched rat tail tendons (see Methods).
Fig.\,4a shows the relative absorption of untreated and pulled collagen fascicles for different buffer incubation times. We detect a significant increase in absorbance at 595\,nm after 30\,min and 60\,min of incubation. In pulled tendons, the hydrogen peroxide concentration is about 1\,$\mathrm{\mu M}$ higher compared to untreated tendons, as evident from reference measurements (Supplementary Fig.\,4a).    
Performing already the tensile loading of the collagen samples  at higher water content, i.e. 50\,\% relative humidity, diminishes the increase in the EPR signal intensity (Supplementary Fig.\,4b). However, the FOX measurements on these samples again detect a pronounced increase in $\mathrm{H}_{2}\mathrm{O}_{2}$ concentration by pulling (Fig.\,4b). Hence, bond scission still occurs homolytically at this humidity, but produced radicals that readily convert to peroxides. In agreement, we immediately detect peroxides already without incubation in wet collagen samples  (Fig.\,4a, 0\,min., compare Fig.\,4b, 0\,min.).

\begin{figure}
\includegraphics[scale=0.21]{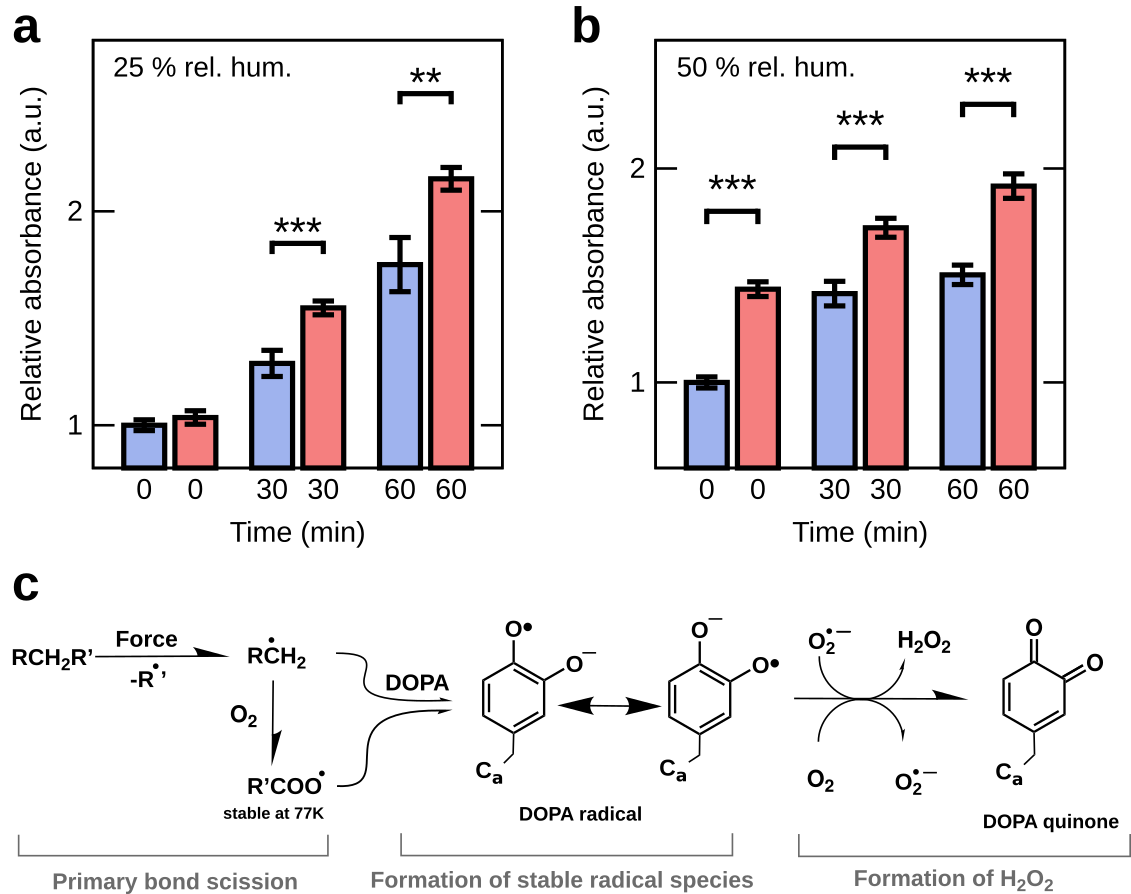}
\caption{\textbf{Collagen produces hydrogen peroxide.} \textbf{a}, Relative absorbance at 595\,nm in the FOX assay for different buffer incubation times. Blue: unstressed collagen fascicles, red: fascicles pulled with 15\,N, both at 25\,\% relative humidity. Bar height shows the mean and error bars the standard error of the mean, calculated based on six measurements from 2 independent sample series (see Methods). \textbf{b}, The same plot as in \textbf{a}, for pulling at 50\,\% relative humidity. Stars in \textbf{a}, and \textbf{b}, indicate statistical significance p-values, $**: p<0.01, ***: p<0.001$. \textbf{c}, Scheme proposing a reaction path from primary bond rupture to hydrogen peroxide formation via DOPA radicals.} 
\end{figure}

\section*{Discussion}

Taken the EPR, MD and QM results together, our data thus suggest a primary irreversible bond rupture in the crosslink regions of collagen I, followed by radical propagation to proximate DOPA.

However heterolytic bond rupture has been shown to be preferred over homolytic bond rupture if the rupturing bond is directly surrounded by water, which then acts as a nucleophile and initiates hydrolysis as opposed to radical formation~\cite{Aktah:2002,Dopieralski:2017,Pill:2014}. Instead, our data show homolytic bond scission to occur to a significant extent in dense collagen tissues even under high humidity, leading to subsequent ROS formation, in close analogy to what has been previously found for hydrated synthetic polymers~\cite{Fitch:2014,Baytekin:2012}. 
Collagen mechanoradicals thus couple subfailure mechanical stress to oxidative stress in tendons and supposedly other connective tissues. DOPA, arising from the -- enzymatic or mechanoradical-mediated -- oxidation of conserved Tyr or Phe residues around collagen crosslink sites, acts as protective radical sink (Fig.\,4c). 

Interestingly, DOPA has been recently observed in other redox-active proteins undergoing electron transfer~\cite{Srinivas:2018, Blaesi:2018}. The observation of DOPA in collagen has also been made previously, with evidence for DOPA-derived crosslinks~\cite{Mcdowell:1999}.

We propose mechanically induced ROS formation as we observe it here to play a role in physical exercise~\cite{Jackson:2016} and redox-mediated pathologies such as pain, inflammation and arthritis.

\newpage
\section*{Methods}

\subsection*{Tensile tests and EPR}\label{sampleprep}
Tendons were cut out of \textit{Sprague Dawley} rat tails from Charles River using a disposable surgery scalpel. Tendon samples were about 2\,cm long with  cross-sectional areas varying along and among samples between 0.5-3\,mm$^2$, and with weights between 20-30\,mg.
After extraction, the collagen tendons were equilibrated at least one hour at room temperature and different humidities.
Tensile tests were performed with LEX810 from Diastron covered by a custom-built humidity chamber to control humidity. Tendons were pulled with forces in the range of 5-20\,N which corresponds to stresses of about 2-40\,MPa depending on the tendon cross-sectional area, 0.5-3.0\,mm$^{2}$. The constant force was reached by increasing the extension within $\sim$200\,s using creep rates of 0.01\,mm/s, and then kept for 1000\,s.\\
For continuous-wave (cw) EPR measurements at X-band ($\sim$9\,GHz frequency), the stressed tendons were put into 3\,mm EPR tubes and measured at room temperature (or liquid nitrogen temperature, Supplementary Fig.\,3b,c) with a Bruker ESP300 setup with standard parameters (2 mW microwave power, 0.2\,mT modulation amplitude, 100\,kHz modulation frequency, 20 scans per sample and load).   
For the real-time measurement, the tendon was attached inside the EPR cavity and pulled at the same time with a load of 350\,g which corresponds to a force of 3.43\,N, using the same EPR setup and parameters.
To prevent radical transfer, we additionally performed both the mechanical loading and the EPR measurement at 77\,K.
For mechanical loading, we crushed the sample in a mortar for 5\,min in liquid nitrogen. We confirmed that crushing in the mortar produces the same type of radicals by crushing and measuring at room temperature and comparing the X-band spectra (Supplementary Fig.\,3a).\\
G-band EPR measurements were performed at 40\,K with a custom-built spectrometer~\cite{Hertel:2005}, upgraded to 100\,mW microwave power on the transmitter output, using a microwave frequency of 180\,GHz and a tendon collagen sample crushed at room temperature. The absorption signal is recorded directly, compared to cw-EPR, where the first derivative is recorded. Pulse mode detection of the EPR signal with a Hahn echo sequence was used for the G-band measurements. The pulse lengths for the $\mathrm{\pi}$/2- and $\mathrm{\pi}$- pulse were 50\,ns and 100\,ns, respectively. The pulse separation time was 250\,ns and experiment repetition time was 10\,ms. The magnetic field sweep rate was 1\,G/s and 100 signals were averaged per magnetic field step. The linearity of the magnetic field was controlled by a MnO in MgO sample as standard reference (see Supplementary Fig.\,3d). 

\subsection*{Collagen fibril modeling}
A full-atom model of collagen fibril from \textit{Rattus norvegicus} was built using an integrative structural modeling approach. In this approach, the overall shape of collagen molecules and their relative packing within collagen fibril was based on the low-resolution fibril structure from fiber diffraction (\textit{Rattus norvegicus}, PDB code: 3HR2~\cite{Orgel:2006}), whereas the local interactions were modeled based on distance restraints derived from high-resolution collagen structures. These restraints were generated based on the collagen model built using THeBuScr program~\cite{Rainey:2004}, which allows building idealized collagen models based on statistical parameters derived from high-resolution collagen-like peptides structures. First, the atomic model of the collagen triple helix was constructed using Modeller~\cite{Vsali:1993} by taking the 3HR2 structure as a template and replacing the template inter-atomic distance restraints with the high-resolution restraints. Then, the model of the “full” collagen fibril was reconstructed by applying the symmetry information derived from 3HR2 structure to the triple helix model. Here, the full model corresponds to the fragment spanning one gap and one overlap region~\cite{Hodge:1963,Orgel:2006} of a bundle of 37 aligned collagen I triple helices (Fig.\,2a) was selected as a representative model. Acetyl and N-methyl groups were attached to the N and C termini of truncated polypeptide chains, respectively. Steric clashes between side chains were resolved using relax protocol in Rosetta~\cite{Nivon:2013,Conway:2014}. To reconstruct covalent HLKNL crosslinks between modified lysine residues first the lysine residues were mutated ~\textit{in silico} to modified residues corresponding to two halves of a crosslink using Modeller, connected by special bonds and followed by energy minimization using GROMACS~\cite{Abraham:2015}. This resulted in the model comprising 12 HLKNL crosslinks (six N-terminal and six C-terminal).

\subsection*{MD simulations}
Structures of crosslink molecules were built using Maestro~\cite{Schrodinger:2016} and geometry optimized with Gaussian09~\cite{Frisch:2009} using the B3LYP functional~\cite{Lee:1988,Becke:1993}. Restrained Electrostatic Potential (RESP) atomic partial charges were calculated using Antechamber~\cite{Wang:2004,Wang:2006}. Parameters were derived using Acpype~\cite{DaSilva:2012}. MD simulations were performed using the GROMACS software~\cite{Abraham:2015}, with the Amber99SB-ildn*~\cite{Best:2009,Lindorff:2010} force field and a time step of 2\,fs. For the simulations, the model of the collagen fibril was solvated in a TIP4P~\cite{Jorgensen:1983} water box of $16\times17.5\times140$\,nm$^{3}$. Then Na\,$\mathrm{^+}$ and Cl\,$\mathrm{^-}$ ions were added at a concentration of 150\,mM. 
An initial energy minimization was performed using the steepest descent method. Thereafter, 10\,ns of NVT followed by 10\,ns of NpT equilibration were performed, both with harmonic positional restraints on the protein heavy atoms of 1000\,kJ/mol/nm. Subsequently, 50\,ns of equilibration with releasing harmonic restraints, followed by 50\,ns unrestrained equilibrium MD simulations were carried out. The temperature was kept constant at 300 K by using a velocity rescaling thermostat~\cite{Bussi:2007} with a coupling time of 0.1 ps. The pressure was kept constant at 1\,bar using isotropic coupling to a Parrinello-Rahman barostat~\cite{Parrinello:1981} with a coupling time of 2.0 ps. In all simulations, the long-range electrostatic interactions were treated with the Particle Mesh Ewald (PME) method~\cite{Darden:1993}. All heavy atom-hydrogen bonds were constrained using the LINCS procedure~\cite{Hess:2008}. A time step of 2\,fs was used.\\
The obtained equilibrated system was subjected to constant force pulling simulation using an average force of 1\,nN per chain, equally and not equally distributed among different triple helices. Additional simulations were carried out with the same average force but unequally distributed into the helices, denoted shear loading (Supplementary Fig.\,2), to take the effect of structural irregularities of real fibrils into account. To mimic the behavior of the longer collagen fibril we prevent unwinding of triple helices at the termini which resulted from the truncation of the system. Torque restraints at the termini of the triple helices were applied using the enforced rotation protocol~\cite{Kutzner:2011}. FDA was used to monitor forces in bonds~\cite{Costescu:2013}. In brief, time-averaged scalar pairwise forces $F_{ij}$ from bond potentials were computed for atom pairs $i,j$ of backbone and crosslinks.

\subsection*{Sequence analysis}
To obtain the multiple sequence alignments (MSAs) of COL1A1 and COL1A2 sequence families for conservation analysis, 3,614 collagen sequences belonging to the collagen orthologous group (KOG3544) were downloaded from the EggNOG database~\cite{Huerta:2015}. The sequences were then clustered using CLANS~\cite{Frickey:2004}. A cluster of sequences containing COL1A1 and COL1A2 and other closely related collagens was extracted and aligned using MAFFT~\cite{Katoh:2002}. From the resulting MSA, a neighbor-joining phylogenetic tree was constructed using JalView and, based on the tree, MSAs comprising COL1A1 (64 sequences) and COL1A2 (59 sequences) were extracted.\\
To assess if the location of redox-active residues in the vicinity of crosslinks is conserved within the two collagen families, a repetitive bundle (five triple helices) of the collagen fibril was selected and divided into slices spanning 10 residues in length along the fibril. The number of the given amino acid type in each slice was calculated by summing up all occurrences of that residue in the sequences within the multiple sequence alignment in the positions of the residues located within the slice.
Protein structures were visualized with UCSF Chimera~\cite{Pettersen:2004}.

\subsection*{QM calculations}
Dimethylamine and major part of hydroxylysino-keto-norleucine (HLKNL) crosslink without the optional hydroxyl group as representatives for collagen backbone and collagen crosslinks were calculated.
First, structures were optimized with constraints on the first and the last heavy atom with Gaussian09~\cite{Frisch:2009} using B3LYP~\cite{Becke:1993} and cc-pVDZ basis set~\cite{Dunning:1989}. The distance between atoms was increased in steps of 0.05\,\AA\, which corresponds to an elongation of every bond by about 5\,\% along an axis. After that, an \textit{ab initio} multireference CASSCF calculation was performed on the optimized structures using Molcas 8~\cite{Aquilante:2016:Molcas} to calculate the bond dissociation energy of the ground state and the first excited state, respectively. The active space contained 6 orbitals populated by 6 electrons. 
EPR g-factors were calculated by unrestricted B3LYP calculations with cc-pVDZ basis set of optimized DOPA radical structures in their protonated and unprotonated forms using ORCA 4.0~\cite{Neese:2018}.

\subsection*{Hydrogen peroxide detection}
Multiple rat tail tendon samples were extracted the same way as described in section \textbf{Tensile test and EPR}. After extraction, the tendons were rinsed, vortexed and centrifuged for 2\,min with 2000\,rpm in PBS buffer without calcium and magnesium chloride and after that dried for 2 hours at RT. The tendon samples were separated in two groups of equal mass, each tendon from the first group was pulled for 200\,s with 15\,N (reached with a creep rate of 0.05\,mm/s) as described in section \textbf{Tensile test and EPR}. Tendons in the second group were kept untreated for reference. The procedure was done at 25\,\% and 50\,\% relative humidities.\\
After treatment the tendons were put into 2\,ml Eppendorf tubes together with 1.7\,ml PBS without calcium chloride and magnesium chloride. After 0\,min, 30\,min and 60\,min, 100\,$\mathrm{\mu}$l supernatant was taken from both, treated and untreated samples, and pipetted in Corning 96 Flat Bottom Transparent Polystyrol well plates. To measure the peroxide concentration, 100\,$\mathrm{\mu}$l working reagent of Pierce$^{TM}$ Quantitative Peroxide Assay Kit from ThermoScientific was added to each well and incubated for 1000\,s in the dark at RT. The oxidation of ferrous to ferric ion in the presence of xylenol orange was used to detect peroxide (FOX). The absorption of xylenol orange was measured with a TECAN infinite M200 Pro at 595\,nm. Per sample and every point in time, three values from different wells were collected.    


\section*{Data availability}
All data that support the results of this study are available from corresponding author upon reasonable request.



\bibliographystyle{naturemag}
\bibliography{radicals}


\subsection*{Acknowledgements}
We are grateful for financial support by the Klaus Tschira Foundation, the Volkswagen Foundation, the Excellence Cluster Cellnetworks and BIOMS of Heidelberg University. The authors acknowledge support by the state of Baden-W{\"u}rttemberg through bwHPC and the German Research Foundation (DFG) through grant INST 35/1134-1 FUGG. We thank the Center of Breath Research of Saarland University Medical Center for donating the rat tails, mechanical workshop of Saarland University Medical Center for technical support and Martin M\"uller of Kiel University for the loan of the LEX810.

\subsection*{Author contributions}
C.Z. performed collagen tensile tests, X-band EPR experiments, QM calculations and hydrogen peroxide detection. A.O.-K. performed collagen fibril modeling, MD simulations and sequence analysis. B.R., D.M. and A.O.-K. supported tensile tests. B.R., U.B. and T.P.D. supported hydrogen peroxide detection. C.Z. and V.D. performed G-band EPR experiments and T.P. and M.B. supported G-band EPR experiments and their interpretation. C.D. supervised QM calculations. R.K. supervised experiments. F.G. supervised the project. F.G., C.Z and A.O.-K. wrote the manuscript. All authors commented on the manuscript and contributed to it.

\subsection*{Competing Intersests}
The authors declare that they have no competing financial interests.

\subsection*{Correspondence}
Correspondence and requests for materials should be addressed to F.G.~(email: frauke.graeter@h-its.org).

\newpage
\section*{Supplementary information for\\ ``Mechanoradicals in tensed tendon collagen as a new source of oxidative stress''\\ including  supplementary figures and table}

\setcounter{table}{0}
\renewcommand{\tablename}{Supplementary Table}
\setcounter{figure}{0}
\renewcommand{\figurename}{Supplementary Figure}

\begin{figure}[H]
\includegraphics[scale=0.15]{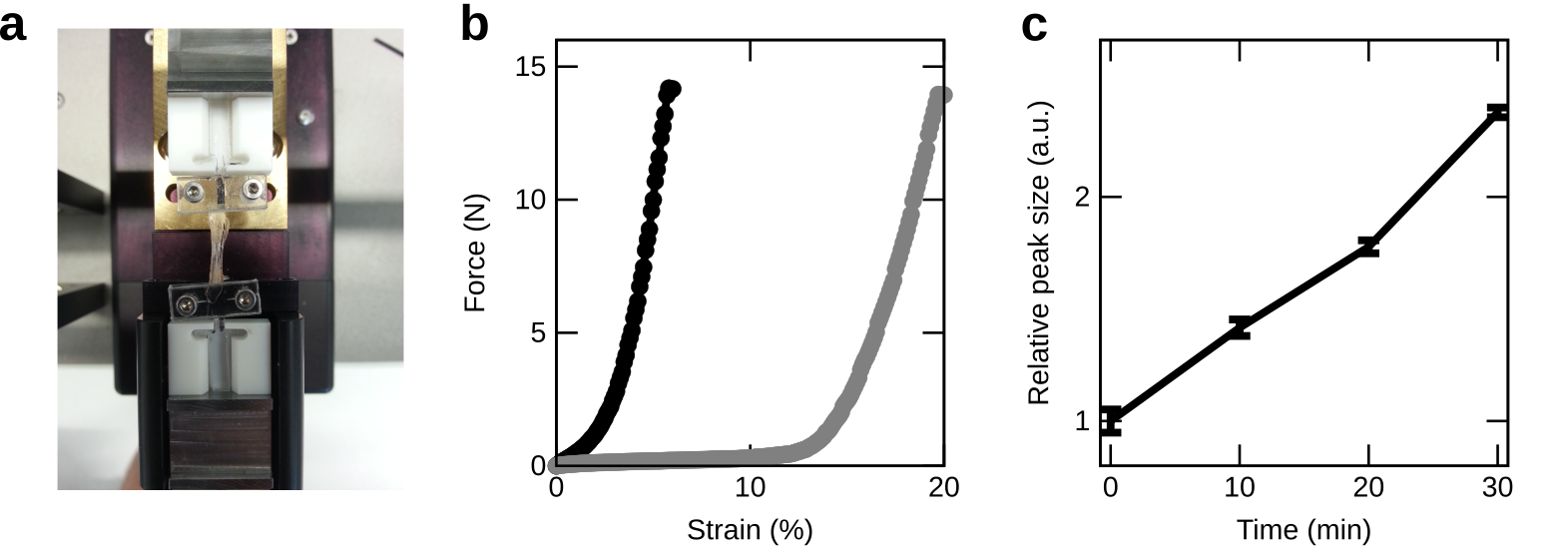}
\caption{\textbf{Strain due to the stress of the extensometer.} \textbf{a}, LEX810 with custom clams to pull on the rat tail tendon piece. \textbf{b}, Stress-strain curve of a 15\,N pulling for 25\,\%hum (black) and 50\,\%hum (gray) without tendon failure. \textbf{c},  Relative EPR peak size recorded at room temperature of a tendon against its pulling time with a load of 200\,g against gravity.}
\end{figure}

\begin{figure}
\includegraphics[scale=0.18]{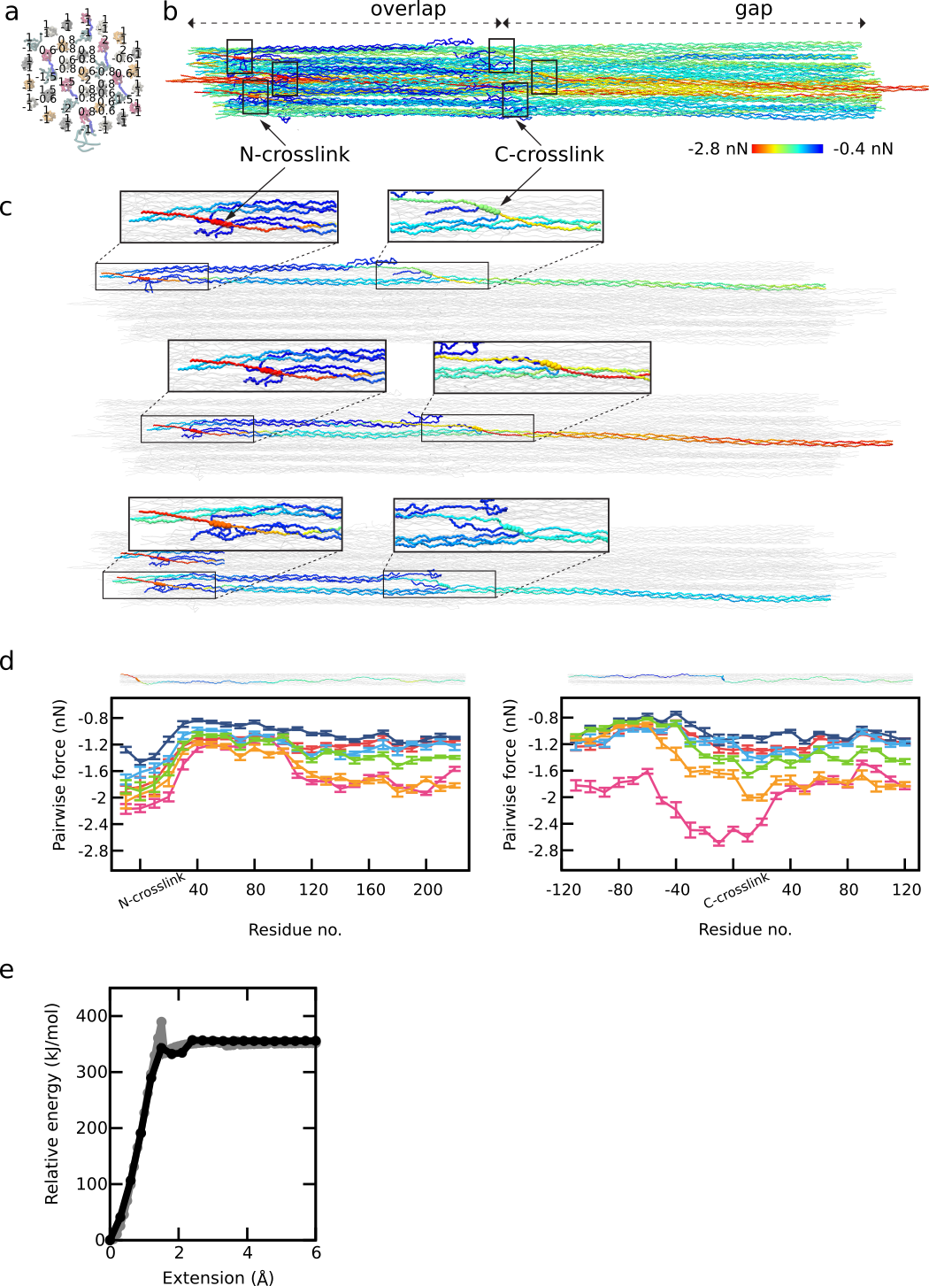}
\caption{\textbf{Force propagation within the collagen fibril during shear pulling simulation.} \textbf{a}, The pulling forces per chain (in nN) are indicated at the cross-section of the fibril. The positive and negative signs designate the pulling direction outward or inwards of the cross-section plane, respectively. \textbf{b}, Random snapshot of the simulation colored according to pairwise force calculated using FDA (blue -- low force, red -- high force). \textbf{c}, Example pairs of overlapping triple helices connected by crosslinks from the snapshot shown in \textbf{b}, depicted separately to better visualize forces around the crosslinks. The example pairs are shown in stick representation colored according to forces as in \textbf{b}, crosslinks are depicted in sphere representation, whereas the remaining collagen chains as gray ribbons. \textbf{d}, Pairwise forces averaged over 10 consecutive residues along the collagen chains connected by crosslinks (left for N-crosslinks, right for C-crosslinks) for all six pairs of overlapping triple helices present in the fibril (with one of the chains shown above the plot aligned to the x-axis). Moving average and standard errors for each curve were calculated over 100 ns simulation using period of 10 ns. \textbf{e}, Bond dissociation energy of carbon-carbon-(black) and carbon-nitrogen bond (gray) in collagen. Relative CASPT2 energies plotted against bond length extension.} 
\end{figure}

\begin{figure}
\includegraphics[scale=0.09]{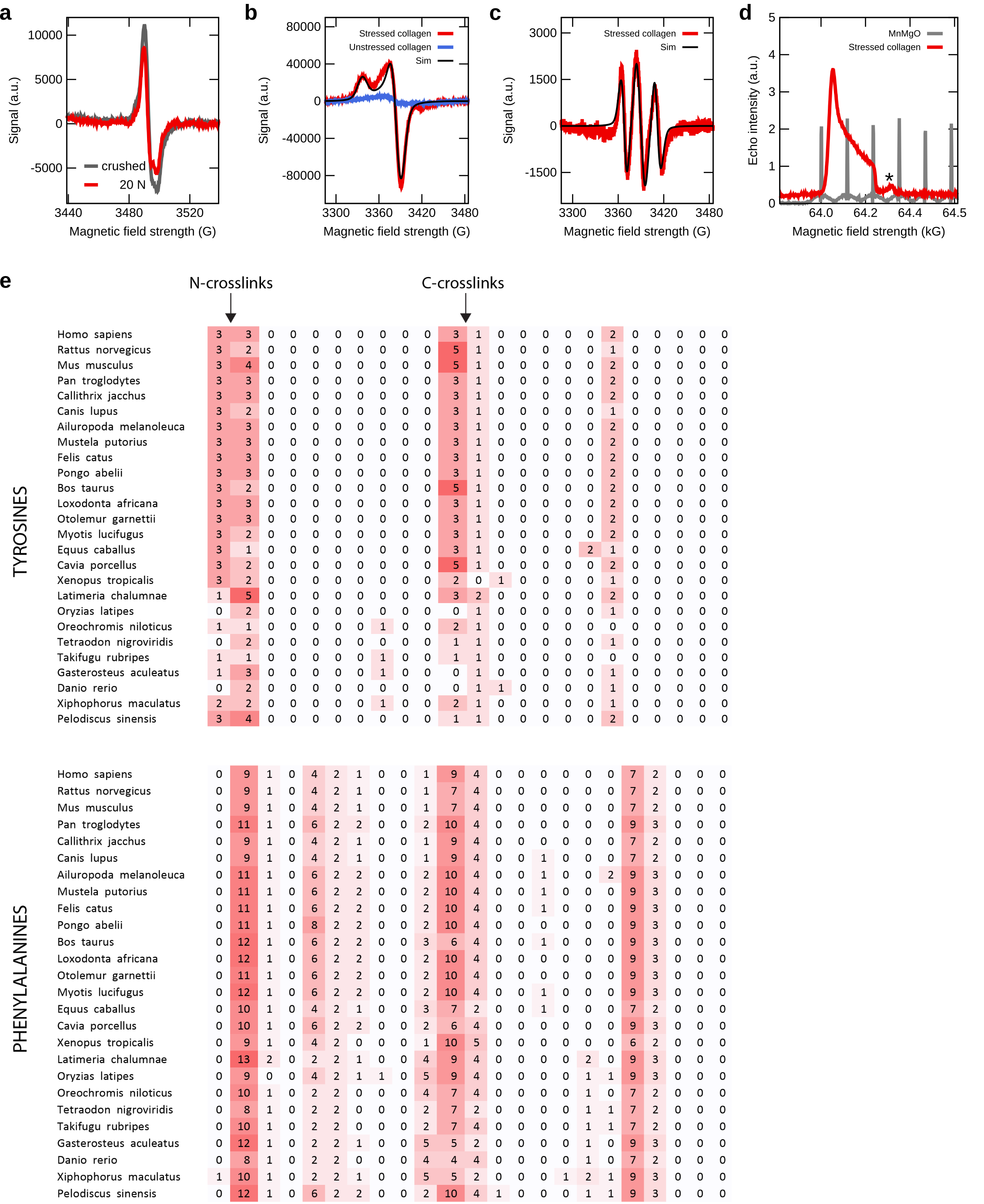} 
\caption{\textbf{a}, Comparison of EPR signals at room temperature of a sample crushed with the mortar for 5\,min (gray) and a sample pulled with 20\,N using the LEX810 (red). The EPR signals are indistinguishable, showing that chemically same molecular scission is at play when pulling and crushing. \textbf{b}, Signal against magnetic field strength for a tendon crushed in liquid nitrogen and measured at 77\,K with a microwave attenuation of 6\,dB is in line with peroxy radical (Supplementary Table\,1). The spectrum of an unstressed tendon at 77\,K and 6\,dB is shown in blue for reference. \textbf{c}, Subtraction of two spectra from the same experiment as in \textbf{b}, 30\,dB and 9\,dB, signal against magnetic field strength suggest a minor contribution from a methylene (CH$_2^*$) radical (Supplementary Table\,1). \textbf{d}, Stressed collagen absorption signal (red) is plotted against the absolute magnetic field. To control the linearity and to determine the absolute magnetic field, MnO in MgO was measured under the same conditions and plotted with a scale factor of 0.1 in gray. The signal marked by * arises from the quartz capillary. \textbf{e}, Counts of Tyr and Phe residues for a given species calculated along the fibril from alignments of representative collagen type I sequences (see Methods). The counts were calculated using a moving average of 10 residues from each chain aligned across the fibril.}
\end{figure}

\begin{figure}
\includegraphics[scale=0.21]{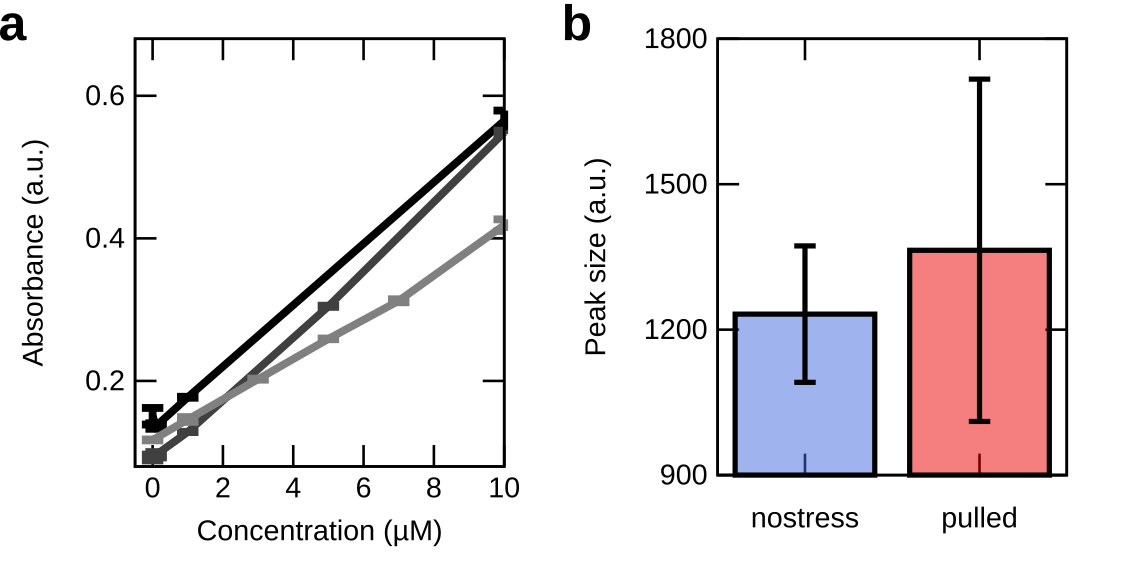}
\caption{\textbf{a}, $\mathrm{H}_{2}\mathrm{O}_{2}$ standard curve of the FOX assay, absorbance at 595\,nm for different $\mathrm{H}_{2}\mathrm{O}_{2}$ concentrations. \textbf{b}, EPR peak size measured at room temperature for tendons kept at 50\,\%hum, in blue unstressed samples and in red samples pulled with a force of 15\,N. Bar height shows the mean and error bars the standard deviation, calculated from 3 independent EPR spectra.}
\end{figure}

\begin{table}
\begin{center}
\caption{EPR parameters for the DOPA, the peroxy and the methylene radical from X-band (X) and G-band (G) measurements at different temperatures and quantum mechanical calculations (QM). DOPA* refers to the radical anion, DOPA-H* to the protonated and neutral radical. Line width and shape values were obtained from X-band or G-band EPR simulations. Line widths and isotropic hyperfine parameters $a_{\mathrm{H}}$ are given in Gauss and temperatures in Kelvin. The number next to hydrogens indicates their position, see Fig.\,3b. Line shape of simulated EPR spectra are defined proportionally by Lorentzian/Gaussian (L/G).}
\begingroup 
\renewcommand{\arraystretch}{1.2} 
\setlength{\tabcolsep}{3pt} 
\begin{tabular}{lllllll}

     & DOPA* X & DOPA* G & DOPA* QM &  DOPA-H* QM & Peroxy* X & Methylene* X \\ \hline
    Temperature & $300$ & $40$ & & & $77$ & $77$ \\ \hline
    $g_{x}$ & $2.0070$ & $2.0071$ & $2.0072$ & $2.0091$ & $2.0346$ & $2.00252$ \\
    $g_{y}$ & $2.0050$ & $2.0066$ & $2.0064$ & $2.0051$ & $2.0065$ & $2.00232$ \\
    $g_{z}$ & $2.0015$ & $2.0022$ & $2.0022$ & $2.0022$ & $2.0013$ & $2.001950$
    \\ \hline
    $a_{\mathrm{H_{\beta1}}}$ & & & $2.8606$ & $0.0912$ &
    \\
    $a_{\mathrm{H_{\beta2}}}$ & & & $0.5492$ & $0.0818$ &
    \\
    $a_{\mathrm{H_{2}}}$ & & & $0.3841$ & $0.4305$ &
    \\
    $a_{\mathrm{H_{5}}}$ & & & $0.3409$ & $0.1677$ &
    \\
    $a_{\mathrm{H_{6}}}$ & & & $0.3890$ & $0.6649$ &
    \\ \hline
    Line width,\\ x,y,z & $3.6, 3.2, 3.0$ & $20.0$ & & & $12.6, 11.0, 12.0$ & $6.0, 9.0, 9.0$ \\
    Line shape & $L/G=1.0$ & $L/G=1.0$ & & & $L/G=0.85$ & $L/G=0.8$ \\
\hline
\end{tabular}
\endgroup
\end{center}
\end{table}

\end{document}